\documentclass[11pt]{article}
\usepackage{amssymb}
\usepackage{cite}
\usepackage{graphicx}
\usepackage{amsmath}
\usepackage{indentfirst}
\usepackage{graphicx}

\begin{document}
\parindent 14pt
\begin{center}
\baselineskip=14pt {\Large\bf Approximate perturbed direct homotopy
reduction method: infinite series reductions to two perturbed mKdV
equations}
\end{center}
 \begin{description}
\item[{}\mbox{}\quad]
\begin{center}
{Xiaoyu Jiao$^a$, Ruoxia Yao$^{a,b,c}$ and S. Y.
Lou$^{a,b,d,}$\footnote{Corresponding Author: S. Y. Lou,
sylou@sjtu.edu.cn}}
\end{center}
\begin{center}
{\small $^a$Department of Physics, Shanghai Jiao Tong
University, Shanghai, 200030,  China\\
$^b$Department of Physics, Ningbo University, Ningbo, 315211,  China\\
$^c$School of Computer Science, Shaanxi Normal University, Xi'an,
710062, China \\
$^d$School of Mathematics, Fudan University, Shanghai, 200433, China
}
\end{center}
{\bf Abstract:} An approximate perturbed direct homotopy reduction
method is proposed and applied to two perturbed modified Korteweg-de
Vries (mKdV) equations with fourth order dispersion and second order
dissipation. The similarity reduction equations are derived to
arbitrary orders. The method is valid not only for single soliton
solution but also for the Painlev\'e II waves and periodic waves
expressed by Jacobi elliptic functions for both fourth order
dispersion and second order dissipation. The method is valid also
for strong perturbations.
 \end{description}

It is very difficult to study nonlinear phenomena lies in the fact
that there are various nonlinear systems which are usually
nonintegrable. For some types of idea cases so-called integrable
models one may use some types of powerful methods (such as the
symmetry reduction method \cite{Olver}, the Darboux transformation
\cite{DT}, the nonlinearization \cite{Cao} or symmetry constraint
method \cite{LiYS} etc) to find some kinds of exact solutions thanks
to there usually exist infinitely many symmetries. However, for real
nonintegrable physical systems, there are only a little of
symmetries or even there is no symmetry at all. In many cases, the
nonintegrable sector of a physical system may company with some
small parameters. In these cases, one may use the perturbation
theory to treat the problems via different approaches. Among these
approaches, the approximate symmetry reduction method may be one of
the best ways \cite{a,b}. To find symmetry reductions, one may use
the classical, nonclassical approaches \cite{Olver,Bluman} and/or
the Clarkson-Kruskal's (CK's) direct method \cite{4,5}. The CK's
direct method is simplest one and can be used to find many group
invariant solutions without using group theory. Furthermore, in more
general cases, the perturbations may not be weak at all. For strong
perturbations, some other types of approaches, such as the homotomy
analysis method (HAM) \cite{Liao} and the Linear \cite{WuLou} and
nonlinear \cite{GaoLou} nonsensitive homotopy approaches etc., have
to be used.

In this letter, we try to combined the CK's direct symmetry
reduction method and HAM to an approximate homotopy direct reduction
approach (AHDRA).

The celebrated modified Korteweg-de Vries (mKdV) equation appears in
many branches of nonlinear science. As one form of approximation,
the singularly perturbed form,
\begin{equation}
u_t+6au^2u_x+u_{x^3}=\epsilon(u_{x^2}+u_{x^4}),\label{1}
\end{equation}
where $a=1$ or $a=-1$, the subscripts $x^n$ mean the
differentiations with respect to $x$ in $n$ times, has arisen in a
number of physical fields, such as models of shallow water on tilted
planes \cite{1}. Soliton perturbation property of the mKdV equation
was analyzed in \cite{2,3,3'}.

In this letter, we consider two special forms of the above equation
\begin{equation}
u_t+6au^2u_x+u_{xxx}=\epsilon u_{x^{3\pm 1}},\label{2}
\end{equation}
with fourth order dispersion (the up sign case) and second order
dissipation (the lower sign case) in terms of APDRA which is a
combination of perturbation theory, direct method and
HAM\cite{4,5,Liao}.

It should be emphasize that in real physical case, the perturbation
terms, say, the parameter $\epsilon$ in \eqref{2}, may not be small.
When the perturbations are not weak, the HAM may be successfully
applied by introducing a homotopy $H(u,q)=0$ of the original model
$A(u)=0$. When the homotopy parameter, $q=0$, the homotopy model
$H_0(u)=H(u,0)=0$ should be solved via known approach. Usually,
$H_0(u)$ is selected as a linear system. In this paper, we select
$H_0(u)$ as an integrable nonlinear system. Concretely, for the
perturbed mKdV system \eqref{2}, we introduce the following linear
homotopy model (linear for the homotopy parameter),
\begin{equation}
(1-q)(u_t+6au^2u_x+u_{xxx})-q(u_t+6au^2u_x+u_{xxx}-\epsilon
u_{x^{3\pm 1}})=0.\label{3}
\end{equation}
It is clear that when $q=0$, \eqref{3} is the well known integrable
mKdV equation which can be solved via many methods. When $q=1$, it
is just the original model \eqref{2}. Now we can solve \eqref{3} via
perturbation approaches by taking  $q$ as a perturbation parameter
no matter $\epsilon $ is small or not.

For Eq. \eqref{3}, according to perturbation theory, the solution
can be expressed as a series of $q$
\begin{equation}
u=\sum_{j=0}^\infty q^ju_j,\label{4}
\end{equation}
with $u_j$ being functions of $x$ and $t$. Substituting Eq.
\eqref{4} into Eq. \eqref{2} and vanishing the coefficients of all
different powers of $q$, we get
\begin{subequations}\label{5}
\begin{eqnarray}
& O(\epsilon^0):& u_{0t}+6au_0^2u_{0x}+u_{0x^3}=0,\\
& O(\epsilon^1):&
u_{1t}+6a(u_0^2u_{1x}+2u_0u_1u_{0x})+u_{1x^3}-\epsilon u_{0x^{3\pm 1}}=0,\\
& O(\epsilon^2):&
u_{2t}+6a(u_0^2u_{2x}+u_1^2u_{0x}+2u_0u_2u_{0x}+2u_0u_1u_{2x})+u_{2x^3}-\epsilon
u_{1x^{3\pm 1}}=0,
\\
&& \cdots~\cdots~\cdots\nonumber\\
& O(\epsilon^j): &
u_{jt}+6a\sum_{k=0}^j\sum_{l=0}^ku_lu_{k-l}u_{j-k,x}+u_{jx^3}-\epsilon u_{j-1,x^{3\pm 1}}=0,\\
&& \cdots~\cdots~\cdots.\nonumber
\end{eqnarray}
\end{subequations}
The similarity solutions for the above equation are of the form
\begin{equation}
u_j=U_j(x,t,P_j(z(x,t))),~(j=0,1,\cdots),\label{6}
\end{equation}
where $U_j$, $P_j$ and $z$ are functions with respect to the
indicated variables and $P_j(z)$ satisfy a system of ordinary
differential equations, which can be obtained by substituting Eq.
\eqref{6} into Eq. \eqref{5}. After the substitution, it is easily
seen that the coefficients for $P_{j,zzz}$ and $P_{j,zz}P_{j,z}$ are
$U_{j,P_j}z_x^3$ and $3aU_{j,P_jP_j}z_x^3$ respectively. We reserve
uppercase Greek letters for undetermined functions of $z$ from now
on. Because the functions $P_j(z)$ dependent only on the variable
$z$, then the ratios of the coefficients are only functions of $z$,
namely,
\begin{equation}
3aU_{j,P_jP_j}z_x^3=U_{j,P_j}z_x^3\Gamma_j(z),~(j=0,1,\cdots),\label{7}
\end{equation}
with the solution $U_j=F_j(x,t)+G_j(x,t){\rm
e}^{\frac{1}{3a}\Gamma_j(z)P_j}\longrightarrow F_j(x,t)+G_j(x,t)
P'_j(z)$. Hence, it is sufficient to seek similarity reductions of
Eq. \eqref{5} in the special form
\begin{equation}
u_j=\alpha_j(x,t)+\beta_j(x,t)P_j(z(x,t)),~(j=0,1,\cdots),\label{8}
\end{equation}
instead of the general form Eq. \eqref{6}.

\noindent\textit{Remark}: Three freedoms in the determination of
$\alpha_j(x,t)$, $\beta_j(x,t)$ and $z(x,t)$ can be notified:

(i) If $\alpha_j(x,t)$ has the form
$\alpha_j(x,t)=\alpha_{j}^\prime(x,t)+\beta_j\Omega(z)$, then one
can take $\Omega(z)=0$;

(ii) If $\beta_j(x,t)$ has the form
$\beta_j(x,t)=\beta_{j}^\prime(x,t)\Omega(z)$, then one can take
$\Omega(z)=$ constant;

(iii) If $z(x,t)$ is determined by $\Omega(z)=z_0(x,t)$, where
$\Omega(z)$ is any invertible function, then one can take
$\Omega(z)=z$.

Substituting Eq. \eqref{8} into Eq. (5a), we can see that the
coefficients for $P_{0zzz}$, $P_{0z}P_0^2$, $P_{0z}P_0$ and
$P_{0zz}$ are $\beta_0z_x^3$, $6a\beta_0^3z_x$,
$12a\alpha_0\beta_0^2z_x$ and $3\beta_{0x}z_x^2+3\beta_0z_xz_{xx}$,
respectively. We require that
\begin{equation}
6a\beta_0^3z_x=\beta_0z_x^3\Psi_0(z),~12a\alpha_0\beta_0^2z_x=\beta_0z_x^3\Phi_0(z),
~3\beta_{0x}z_x^2+3\beta_0z_xz_{xx}=\beta_0z_x^3\Omega_0(z),\label{9}
\end{equation}
and thus, applying \textit{Remark} (i), (ii) and (iii), we have
\begin{equation}
\alpha_0(x,t)=0,~\beta_0(x,t)=z_x,~z(x,t)=\theta(t)x+\sigma(t).\label{10}
\end{equation}
Eq. (5a) is then simplified to
\begin{equation}
\theta^4P_{0zzz}+6\theta^4P_0^2P_{0z}+(\theta_t
z-\theta_t\sigma+\theta\sigma_t)P_{0z}+\theta_t P_0=0.\label{11}
\end{equation}
From the coefficients of $P_{0zzz}$, $zP_{0z}$ and $P_{0z}$ of the
above equation, it is easily seen that
\begin{equation}
\theta_t=A\theta^4,~-\theta_t\sigma+\theta\sigma_t=B\theta^4,\label{12}
\end{equation}
with $A$ and $B$ being arbitrary constants.

When $A\neq0$, Eq. \eqref{12} has the solution
\begin{equation}
\theta=-(3A(t-t_0))^{-\frac{1}{3}},~\sigma=-\frac{B}{A}+s_0(t-t_0)^{-\frac{1}{3}},\label{13}
\end{equation}
where $s_0$ and $t_0$ are arbitrary constants.

On substitution of Eq. \eqref{8} into the general form Eq. (5d), the
coefficients of $P_{j,z^3}$, $P_{j-1,z^{3\pm1}}$ and $P_{0z}P_0$ are
$\beta_jz_x^3$, $\mp \epsilon \beta_{j-1}z_x^4$ and
$12a\alpha_j\beta_0^2z_x$, respectively, which lead to
\begin{equation}
\beta_{j-1}z_x^4=\mp \epsilon
\beta_jz_x^3\Psi_j(z),~12a\alpha_j\beta_0^2z_x=\beta_jz_x^3\Phi_j(z),~(j=1,2,\cdots).\label{14}
\end{equation}
By $\beta_0=z_x$, \textit{Remark} (i) and (ii), we obtain
\begin{equation}
\alpha_j=0,~\beta_j=z_x^{1\pm j},~(j=0,1,\cdots).\label{15}
\end{equation}
Eq. \eqref{4}, \eqref{8}, \eqref{10}, \eqref{13} and \eqref{15}
determine the perturbation series solution to Eq. \eqref{2}
\begin{equation}
u=\sum_{j=0}^\infty(-1)^{j+1}(3A(t-t_0))^{-\frac{1}{3}(1\pm
j)}\epsilon^jP_j(z),\label{16}
\end{equation}
with the similarity variable
$z=-x(3A(t-t_0))^{-\frac{1}{3}}+s_0(t-t_0)^{-\frac{1}{3}}-\frac{B}{A}$
and the similarity reduction equations are
\begin{equation}
P_{j,z^3}=-(j+1)AP_j-(Az+B)P_{j,z}-6a\sum_{k=0}^j\sum_{l=0}^kP_lP_{k-l}P_{j-k,z}+\epsilon
P_{j-1,z^{3\pm1}},~(j=0,1,\cdots),\label{17}
\end{equation}
with $P_{-1}=0$. When $j=0$, Eq. \eqref{17} degenerates to
Painlev\'e II type equation.

When $A=0$, Eq. \eqref{12} has the solution
\begin{equation}
\theta=t_0,~\sigma=Bt_0^3t+s_0,\label{18}
\end{equation}
where $s_0$ and $t_0$ are arbitrary constants. From Eq.
\eqref{10}, Eq. \eqref{18} implies an equivalent travelling wave
form $z=x+ct$, so that we obtain the perturbation series
travelling wave solution to Eq. \eqref{2}
\begin{equation}
u=\sum_{j=0}^\infty\epsilon^jP_j(z),\qquad z=x+ct,\label{19}
\end{equation}
where all $P_j(z)$ satisfy
\begin{equation}
cP_j+2a\sum_{k=0}^j\sum_{l=0}^kP_{j-k}P_{k-l}P_l+P_{j,z^2}-\epsilon
P_{j-1,z^{2\pm1}}+a_j=0,~(j=0,1,\cdots),\label{20}
\end{equation}
with $P_{-1}=0$.

Taking $j=0$, it is obvious that Eq. \eqref{20} becomes the zeroth
order equation, the well known mKdV equation which has the general
solution
\begin{equation}
\int^{P_0}\frac{{\rm d} p}{\sqrt{c_0-ap^4-cp^2-2a_0p}}=\pm
(z-z_0)\label{20'}
\end{equation}
with arbitrary constants $c,\ a_0, c_0$ and $z_0$. It is also
interesting that for the series travelling wave solution, it is
not difficult to find the solution of $P_j$ can be expressed by
$P_0$, result reads
\begin{eqnarray}
P_j&=&\sqrt{c_0-2a_0P_0-cP_0^2-aP_0^4}\left[c_{1j}+c_{2j}\int^{P_0}\frac{{\rm
d} p}{\sqrt{(c_0-ap^4-cp^2-2a_0p)^3}}\right.\nonumber\\
&&\left.-\int^{P_0}\int^{p'}
\frac{F_j(p)}{\sqrt{(c_0-ap'^4-cp'^2-2a_0p')^3}} {\rm d} p{\rm d}
p'\right],\label{20"}
\end{eqnarray}
where $c_{1j}$ and $c_{2j}$ are arbitrary constants while
\begin{equation}
F_j(P_0)\equiv
2a\sum_{k=1}^{j-1}\sum_{l=0}^kP_{j-k}P_{k-l}P_l+2aP_{0}\sum_{l=1}^{j-1}P_{j-l}P_l-\epsilon
P_{j-1,z^{3\pm1}}+a_j.
\end{equation}

The general solution \eqref{20'} can be rewritten as some types of
Jacobi elliptic functions \cite{LouTang}. For some special
selections of the constants, it can be written as some types of
soliton solutions or periodic wave solutions, for instance, if we
select $a=-1,\ c=2k^2,\ a_0=0,\ c_0=k^4$ for the up sign (the
dissipative case), then we have the hyperbolic tangent shape kink
soliton solution
\begin{equation}
P_0= k{\rm tanh}(kx+2k^3t)\equiv kT.\label{21}
\end{equation}

\textit{Remark:} The convergence of infinite series solution Eq.
\eqref{16} is superior to the fourth order dispersion case (the up
sign case), because the general terms of Eq. \eqref{16} become
infinitesimal for sufficiently large time $t$,
$$|3A(t-t_0)|\gg1. $$
For the infinite series solution \eqref{16} with the lower sign (the
dissipative case), the series will be convergent for not very large
time, i.e. for
$$|3A(t-t_0)|\ll1. $$

More specifically, for the dark solitary wave solution \eqref{21},
we can easily find the closed forms for the higher order homotopy
perturbation solutions. Here is a explicit form of the eighth order
approximate solution (the end point condition $q=1$ has been used)
\begin{eqnarray}
u&=&kT+\frac{\epsilon}6+\frac{S^2T-2T}{48k}\epsilon^2-\frac{2T+BS^2(BT+1)}{2034k^3}\epsilon^4
-\frac{12T+3B^3S^4-2B(B^2-3)S^2}{331776k^5}\epsilon^6\nonumber\\
&&+\frac{3B^3S^4(BT-1)-BS^2(TB^3+15-3BT-2B^2)-30T}{15925248k^7}\epsilon^8+\cdots
,\label{21}
\end{eqnarray}
where
$$S\equiv \sqrt{1-T^2},\ B\equiv \ln\frac{1-T}{1+T}.$$
\input epsf
\begin{figure}
\centerline{\epsfxsize=6cm\epsfysize=5cm\epsfbox{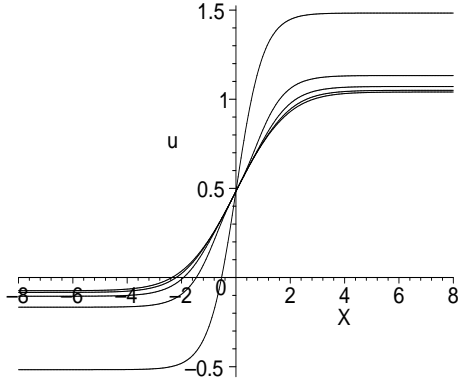}}
\caption{The plots of the perturbed kink solitary wave solution for
the orders 1,\ 2,\ 4,\ 6 and 8 respectively from upper to lower of
the right side of the figure while the parameters are fixed as
$\epsilon=2.9,\ k=1$.}
\end{figure}
It should be emphasized that homotopy approximation convergence
quite well not only for weak perturbation (small $\epsilon$) but
also for strong perturbations. Fig. 1 shows the schematic plots of
the first five approximants with respect to the orders 1,\ 2,\ 4,\ 6
and 8 respectively from upper to lower of the right side of the
figure while the parameters are fixed as
\begin{eqnarray}
\epsilon=2.9,\ k=1. \label{21}
\end{eqnarray}
 From the figure, we find that the lines of the sixth order and the
 eighth order are almost stuck together though the ``perturbed
 parameter" $\epsilon=2.9$ which is not a small one!

Similar to the HAM, the APDRA is applicable to other perturbed
nonlinear partial differential equations with and without small
parameters and it is thought-provoking to explore a general
principle for the perturbed nonlinear partial differential equations
holding similar results. Different from the HAM, we take the zeroth
order as an nonlinear integrable system instead of a linear one,
which largely modified the convergence rate. Here we take the direct
method as a tool to find the approximate symmetries and symmetry
reductions. The similar results can also be obtained via approximate
classical and nonclassical symmetry reduction approaches which may
be used to the KdV-Burgers equation \cite{b}, the perturbed
nonlinear Schr\"odinger systems \cite{JIa} and the perturbed
Boussinesq system\cite{Yao}.

\noindent{\bf Acknowledgement}

The work was supported by the National Natural Science Foundations
of China (Nos. 10735030, 10475055, 10675065 and 90503006), National
Basic Research Program of China (973 Program 2007CB814800) and
PCSIRT (IRT0734), and the Research Fund of Postdoctoral of China
(No.20070410727).

\end{document}